\documentclass[final,5p,times,twocolumn]{elsarticle}
\usepackage{amsbsy}
\usepackage{amssymb}
\usepackage{amsmath}
\newcommand{\beq}{\begin{equation}}
\newcommand{\eeq}{\end{equation}}
\newcommand{\be}{\begin{eqnarray}}
\newcommand{\ee}{\end{eqnarray}}
\journal{Astroparticle Physics}
\usepackage{graphicx}
\usepackage{slashed}
\usepackage{amsmath}
\usepackage{amssymb}
\usepackage{txfonts}               
\usepackage{longtable,lscape}
\usepackage{hyperref}
\usepackage{tabu}
\newcommand{\bemul}{\begin{multline}}
\newcommand{\eemul}{\end{multline}}
\begin{document}
\begin{frontmatter}
\title{On the possibilities of high-energy neutrino production in the jets of microquasar SS433 in light of new observational data}
\author[label1]{Mat\'{\i}as M. Reynoso}
\author[label1]{Agust\'{\i}n M. Carulli}
\address[label1]{IFIMAR (CONICET-UNMdP) and Departamento de F\'{\i}sica, Facultad de Ciencias Exactas y Naturales, Universidad Nacional de Mar del Plata, Funes 3350, (7600) Mar del Plata, Argentina}
%
\begin{abstract}
{Microquasar SS433 is composed by a supergiant star that feeds mass through a supercritical accretion disk to a  $\sim 10 \ M_\odot$ black hole. The latter launches two oppositely directed jets that precess with a period of $162$ days. The system has been detected at different spatial scales in frequencies ranging from radio to gamma rays, and has long been considered as a potential neutrino source which has been observed by AMANDA in the past, and later IceCube, leading to more restrictive upper bounds on the neutrino flux. }
{ In this work, we explore the possibilities that neutrinos could be produced in the jets of this source at levels consistent, or at least, not incompatible with any current data on electromagnetic emission available.}
{ In order to do so, we consider the injection of both electrons and protons at different positions in the jets, and we compute their broadband photon emission by synchrotron and interactions with ambient photons and matter. After correcting the high energy photon flux by the effect of $\gamma\gamma$ and $\gamma N$ absorption, we obtain the surviving flux that arrive on Earth and compare it with observational data by gamma-ray detectors.} 
{The flux of high energy neutrinos is consistently computed and we find that if they are eventually detected with IceCube, production must take place at the inner jets, where gamma-ray absorption is important, in order to avoid current VHE constraints form HESS and MAGIC. {Additionally, we find that if the flux of 25 TeV gamma-rays recently detected with HAWC and which corresponds to the jet termination region were produced mainly by $pp$ interactions, this would lead to a too faint neutrino flux that is beyond the reach of IceCube in its present configuration.}}
{}
\end{abstract}
\end{frontmatter}
\section{Introduction}

{Microquasars are galactic binary systems composed by a star orbiting around a compact object which launches jets that present non-thermal emission \cite{mirabel1998}. These type of sources have been detected throughout all the electromagnetic spectrum, from radio to even very high energy (VHE) gamma-ray emission \citep{bosch2008}. The particular case of SS433 is unique, since its central black hole (BH) of mass $M_{\rm bh}\simeq 10 \, M_\odot$ launches two powerful and persistent jets  ($L_{\rm k}=10^{39}{\rm erg \ s^{-1}}$) which move in precession with a period of $162$ days (for a review, see Ref. \citep{fabrika2006}). The accretion disk is in a supercritical regime which also moves in precession, and presents a wind that emits photons in microwave and UV wavelengths \citep{fuchs2006, gies2002}. The presence of baryons in the jets has been revealed through the detection of Fe lines \citep{migliari2002}. Radio emission has been detected from the jets at up to large scale where they are stopped by the interstellar medium \citep{dubner1998}, and also through bolides in jets \citep{jeffrey2016}. X-rays are detected presenting orbital and precessional modulation, indicating that they can be produced in the inner jets or in a corona around the compact object \citep{chere2005}.
As for gamma rays, Fermi LAT has detected emission for energies $\sim (1-10) {\rm MeV}$ \citep{bordas2015}, which does not present any modulation due to $\gamma \gamma$ or $\gamma N$ absorption as it would be expected if the emission was generated at the inner regions of the system ($z<10^{12}{\rm cm}$). {This is consistent with a production zone placed at the terminal regions of the jets, where they interact with the external medium W50 \citep{xing2018}}. VHE gamma-ray observations have been long performed with different instruments. A joint search by HESS+MAGIC has led to very stringent constraints for gamma-rays in the energy range $E_\gamma \sim (200-5000)\, \rm{GeV}$ emitted from SS433 \citep{hessmagic2017}, while VERITAS has also produced updated upper limits \citep{veritas2017}. {At higher energies ($E_\gamma\sim 25$ TeV), VHE photons were recenlty detected by HAWC, and the emission was found to be produced at the terminal regions of jets, at distances $\sim 10^{19-20}\rm{cm}$ from the central BH \citep{hawc2018}.} 

As it was pointed out in Ref. \cite{reynoso2008b}, a connection can be established between the gamma-ray and the neutrino fluxes produced by hadronic and photo-hadronic processes, and, in order to establish such link, it is crucial to correct the emitted gamma rays by the effects of absorption in the system that are significant at the inner regions of the jets. In the present work, we take into account such effects, and making use of a simple one-zone model, we compute the gamma-ray and neutrino output that can be expected under different assumptions on the location of the emission zone, the energy dependence, and the total power of the primary electrons and protons injected in such region. Taking into account the last upper limit by IceCube \citep{icecube2017}, we obtain allowed neutrino fluxes and we compare the corresponding co-produced photon fluxes with available multiwavelength observational data and constraints. In particular, the possibility that protons were accelerated in the jet at distances $z_{\rm acc}\gtrsim  5 \times 10^{12}{\rm cm}$ from the central black hole and with a spectrum $\sim E_p^{-2}$ is bounded by the HESS+MAGIC limit \citep{hessmagic2017}; gamma-ray and neutrino production cannot take place at such distances because unabsorbed VHE gamma rays would have been observed.{ Additionally, the neutrino flux that could be expected from the terminal regions of the jet would be too faint to be detected by IceCube.} }

This work is organized as follows. In Section 2, we describe the model employed, the basic physical processes, and we obtain the particle distributions in the jet for different parameter sets. In Section 3, we compute the corresponding output of broadband photons and neutrinos, and compare it with experimental observations. Finally, we present a discussion with our concluding remarks in Section 4.

\section{Basic scenario and physical processes}
The model adopted in the present work is based on previous ones used to describe the broadband emission in microquasars \citep{romero2003,romerovila2008, magnetic2009, vilaromero2010}.
The jets are modelled as cones with a half opening angle $\xi= 0.6^\circ$, so that the radius at a distance
from the black hole $z_0=500 \, R_{\rm Sch}\simeq 1.3\times 10^9$ cm is $R_{\rm jet}= z_0 \tan \xi$. Assuming that equipartition between kinetic and magnetic energy holds at $z_0$,
$B_0=\sqrt{8\pi \rho_{\rm k}}$, where $\rho_{\rm k}=\frac{L_{\rm k}}{\Gamma(\Gamma-1)m_p c^2 \pi \, R_{\rm jet}^2}$.
If the magnetic field drops with the distance as $B(z)=B_0 (z_0/z)^{m}$, with $m\in (1,2)$ \cite{krolik1999}, then sub-partition takes place for $z>z_0$, which favors shock formation by the collision of plasma outflows with different velocities \citep{gaisser1991}. Primary particles (electrons and protons) can be shock-accelerated by the Fermi's first order mechanism \citep{rieger2007}. Still, our main conclusions will remain qualitatively valid regardless of the particular acceleration mechanism at work, as long as it produces a power-law injection of particles as the ones we shall consider below. At the inner jet, we place the injection zone at a distance $z_{\rm acc}=10^{10-13}${cm} with a size $\Delta z$ along the jet, where primary electrons and protons are injected according to
\be
 Q_{i}(E)= K_i E^{-\alpha} e^{-\frac{E}{E_{i,{\rm max}}}} \  \ [\rm  ({GeV} \ cm^3\ sr\ s)^{-1}] \nonumber.
\ee
Here, $\alpha$ is the index of the power law in energy of the particles, $E_{i,{\rm max}}$ is the maximum energy fixed by the balance between an acceleration rate \citep{begelman1990} $t_{\rm acc}^{-1}= \eta {eB}/{E_i}$\footnote{{The value coefficient of acceleration efficiency $\eta$ depends on the specific mechanism at work, and it is, in particular, a factor $(v_{\rm s}/V_A)^2$ slower for the Fermi II stochastic acceleration mechanism as compared to the Fermi I shock acceleration one, for a shock velocity $v_{\rm s}$ and Alfven velocity $V_A$ \citep{rieger2007}.}}  and the
total energy loss rate $t^{-1}_{\rm loss}$. { This balance energy is taken as the maximum energy if it satisfies the Hillas criterion \citep{hillas1984}, i.e., $E_{i,{\rm max}}<E_{\rm H}= e B R_{\rm j}$, and otherwise, we adopt $E_{i,{\rm max}}= E_{\rm H}$.} 

The constants $K_i$ are fixed by normalization on the total power injected in electrons and protons,
\be
L_i= 4\pi \Delta V \int_{E_{i,\rm min}}^\infty dE E\, Q_{i}(E),
\ee
where $E_{i,\rm min}$ is the minimum energy of injection and $\Delta V$ is the volume of the zone.{ In the case of the terminal regions of the jet, we assume that they are located at $z_{\rm acc}\sim 5\times 10^{19}{\rm}$, in line with observational data of HAWC, and we adopt a magnetic field in the range $(10^{-5}-10^{-4}){\rm G}$ \citep{bordas2009}.}

We solve the following transport equation in a steady state for the particle distributions in the jet co-moving frame, $N_i$   with $i=\lbrace e,p,\pi^\pm, \mu^\pm \rbrace$,
 \be
\frac{d\left[b_{i}N_i\right]}{dE_i}+ \frac{N_i}{T_{\rm esc}}= Q_i.\label{equ},
 \ee
where $b_i(E)= -E \ t^{-1}_{\rm loss}(E)$ gives the energy loss for each particle type, $T_{\rm esc}$ is the escape timescale, which is given by $T_{\rm esc}= \Delta z/v_{\rm jet}$ {at the inner jet \citep{romerovila2008, vilaromero2010}, and by $T_{\rm esc}=  R_{\rm j}^2/(2D_{\rm d})$ for an emission zone placed at a terminal region of the jet \citep{bordas2015}. Here, $D_{\rm d	}=D_0(E/{\rm GeV})^{\delta}$ is the diffusion coefficient,  with typical values $D_0\sim 10^{28}{\rm cm^2s^{-1}}$ and $\delta\sim(0.3-0.6)$  \citep{venya1990}, although much smaller values of $D_0$ can not be ruled out, especially for dense regions \citep{ormes1988,hawc2018}.} 

\begin{figure*}[tbp]
\includegraphics[width=.99\textwidth,trim=0 0 0 0,clip]{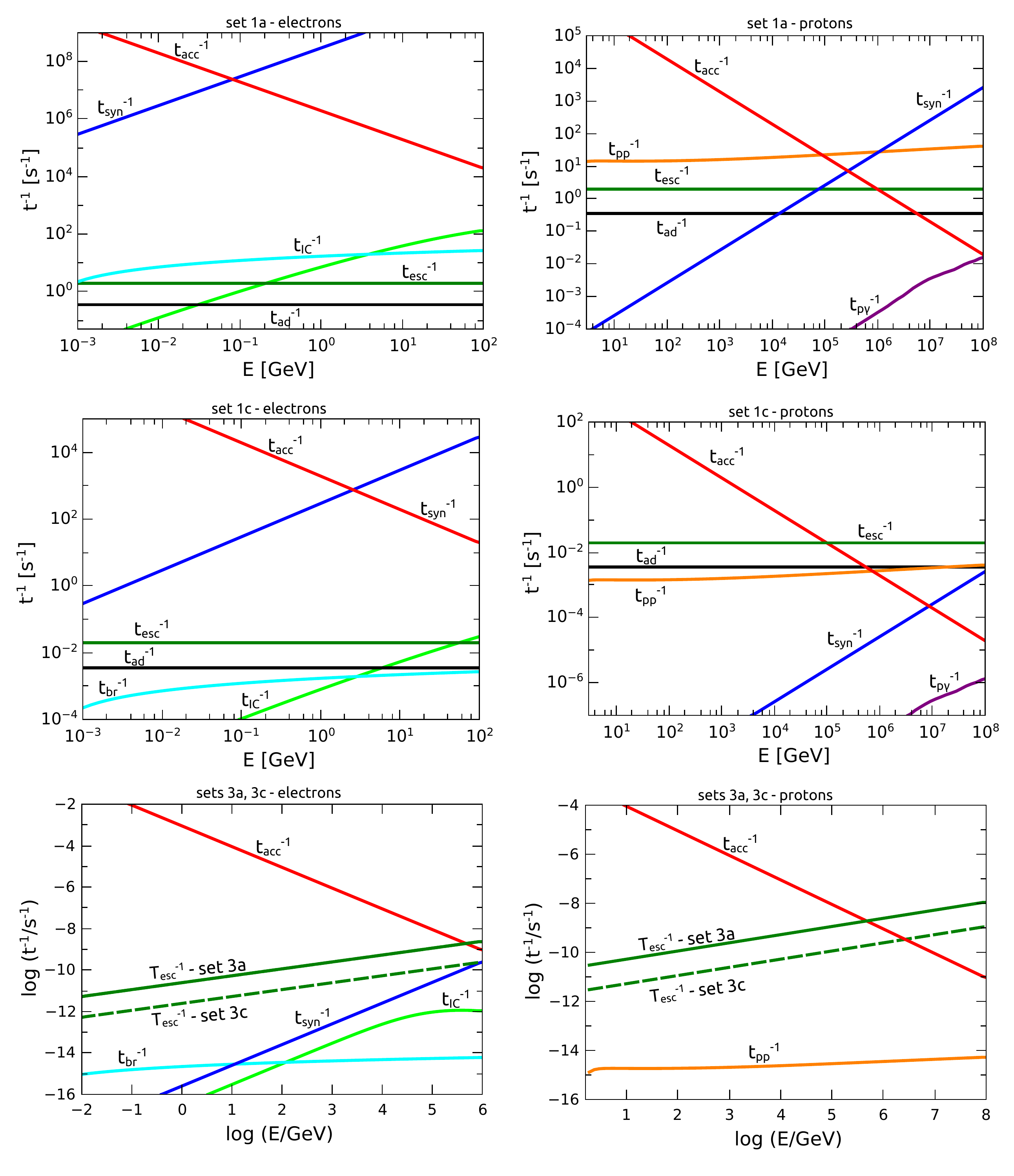}
\caption{Acceleration and cooling rates for electrons (left plots) and protons (right plots) for the parameter values of sets 1a and 1c (top panels), 2a and 2c (middle panels) and 3a and 3c (botom panels), as listed on Table \ref{table.params}. \label{fig:eprates}}
\end{figure*}

In the case of electrons, the cooling processes considered are, in principle, synchrotron emission, {bremsstrhalung}, inverse Compton (IC) interactions with the synchrotron photons at  the inner jet (synchrotron self-Compton, SSC), {and IC interactions with the Cosmic Microwave Background (CMB) in the case of the terminal jet regions}. As for protons, the cooling processes are $pp$ collisions, adiabatic losses by lateral expansion of the jet, synchrotron emission and $p\gamma$ interactions. We obtain these rates following, e.g., Refs. \citep{romerovila2008, leptohadronic2011}, and we show in Fig. \ref{fig:eprates} the obtained rates for electrons and protons under different sets of parameters, which are listed in Table \ref{table.params}.

\begin{table*}
\begin{center}
\small{
\begin{tabular}{|c|c|c|c|c|c|c|} 
\hline 
{Fixed parameters} & \multicolumn{6}{|c|}{values}  \\ 
\hline 
    $L_{\rm k} $: jet kinetic power & \multicolumn{6}{|c|}{$10^{39}{\rm erg \ s}^{-1}$}\\
    $v_{\rm j} $: jet velocity  & \multicolumn{6}{|c|}{$0.26 \,c$} \\
    $M_{\rm bh} $: black hole mass & \multicolumn{6}{|c|}{$10 \ M_\odot$}\\
    $i_{\rm j}$: line of sight angle & \multicolumn{6}{|c|}{$57^\circ$}\\
    $m $: index for $B$ variation & \multicolumn{6}{|c|}{$1.5$} \\
\hline    
{Parameters (sets 1,2)}  & set 1a & set 1b & set 1c & set 2a & set 2b & set 2c \\ 
\hline 
    {$\alpha$: power-law index}  & \multicolumn{3}{|c|}{1.7} &\multicolumn{3}{|c|}{2} \\
    \hline
  $q_{\rm m}$: $\frac{{\rm magnetic \ energy}}{\rm kinetic \ energy}$& $0.1$ & $0.01$ & $0.001$ & \multicolumn{3}{|c|}{0.001} \\
  \hline
$ \eta$: acceleration efficiency   & \multicolumn{3}{|c|}{$2\times 10^{-5}$} & $ 10^{-2}$ &
$10^{-3}$ & $10^{-4}$ \\
\hline
 $\Delta z$: inj. zone thickness [cm]  & $1.5\times 10^{10}$ & $1.5\times 10^{11}$& $1.5\times 10^{12}$ &\multicolumn{3}{|c|}{$1.5\times 10^{13}$}\\
 \hline 
 $q_{\rm rel}$: ratio $(L_e+L_p)/L_{\rm k}$ &  \multicolumn{3}{|c|}{$2\times 10^{-4}$} & \multicolumn{3}{|c|}{$5\times 10^{-4}$}  \\
 \hline
 $a$: ratio $L_p/L_{e}$ &  \multicolumn{3}{|c|}{1}& \multicolumn{3}{|c|}{10} \\
 \hline
{Derived params. (sets 1,2)}  & set 1a & set 1b & set 1c & set 2a & set 2b & set 2c \\ 
\hline 
   $z_{\rm acc}$: inj. zone posiition [cm]&  $1.5\times 10^{10}$ & $1.5\times 10^{11}$& $1.5\times 10^{12}$ & \multicolumn{3}{|c|}{$1.5\times 10^{12}$}   \\
   \hline
    $B$: magnetic field [G] & $  10^6$ &$3.4\times 10^5$ & $10^5$ & \multicolumn{3}{|c|}{$10^5$}\\
    \hline
  $E_{e,{\rm max}}$: Max. $e$ energy  [GeV] &  $0.8$ & $0.5$ & $2.5$ & $57$ & $18$ & $5.7$    \\
  \hline
  $E_{p,{\rm max}}$: Max. $p$ energy  [GeV] &  $8\times 10^4$ & $2.2\times 10^5$ & $3.2\times 10^5$ & $5\times 10^7$ & $1.3\times 10^7$ & $1.5\times 10^6$   \\
       \hline\hline
{Parameters (sets 3,4)}  & set 3a & set 3b & set 3c & set 4a & set 4b & set 4c \\ 
\hline
    {$\alpha$: power-law index}  & \multicolumn{3}{|c|}{1.9} &\multicolumn{3}{|c|}{1.7} \\
    \hline
  $B$: magnetic field [G]& \multicolumn{3}{|c|}{$10^{-5}$} & $4\times 10^{-5} $ & $6\times 10^{-5}$ &$10^{-4} $ \\
\hline 
  $D_{\rm d}$: diffusion coeff. [${\rm cm}^2s^{-1}$]& $3\times 10^{27}$ & $10^{28}$ & $3\times 10^{28}$ & \multicolumn{3}{|c|}{$5\times 10^{26}$} \\
\hline 
  $\delta$: diffusion power & \multicolumn{6}{|c|}{$1/3$} \\
\hline 
$ \eta$: acceleration efficiency   & \multicolumn{6}{|c|}{$10^{-2}$}  \\
  \hline
 $\Delta z$: inj. zone thickness [cm]  & \multicolumn{6}{|c|}{$ 5\times 10^{19} $}\\
 \hline 
 $q_{\rm rel}$: ratio $(L_e+L_p)/L_{\rm k}$ &  \multicolumn{6}{|c|}{$0.1$}  \\
 \hline
 $a$: ratio $L_p/L_{e}$ &  \multicolumn{3}{|c|}{100}& \multicolumn{3}{|c|}{500} \\
 \hline
\end{tabular}}
\end{center}
\caption{Parameters of the model and the different sets of values which are adopted in this work.} \label{table.params}
\end{table*}

\begin{figure*}[tbp]
\includegraphics[width=.99\textwidth,trim=0 0 0 0,clip]{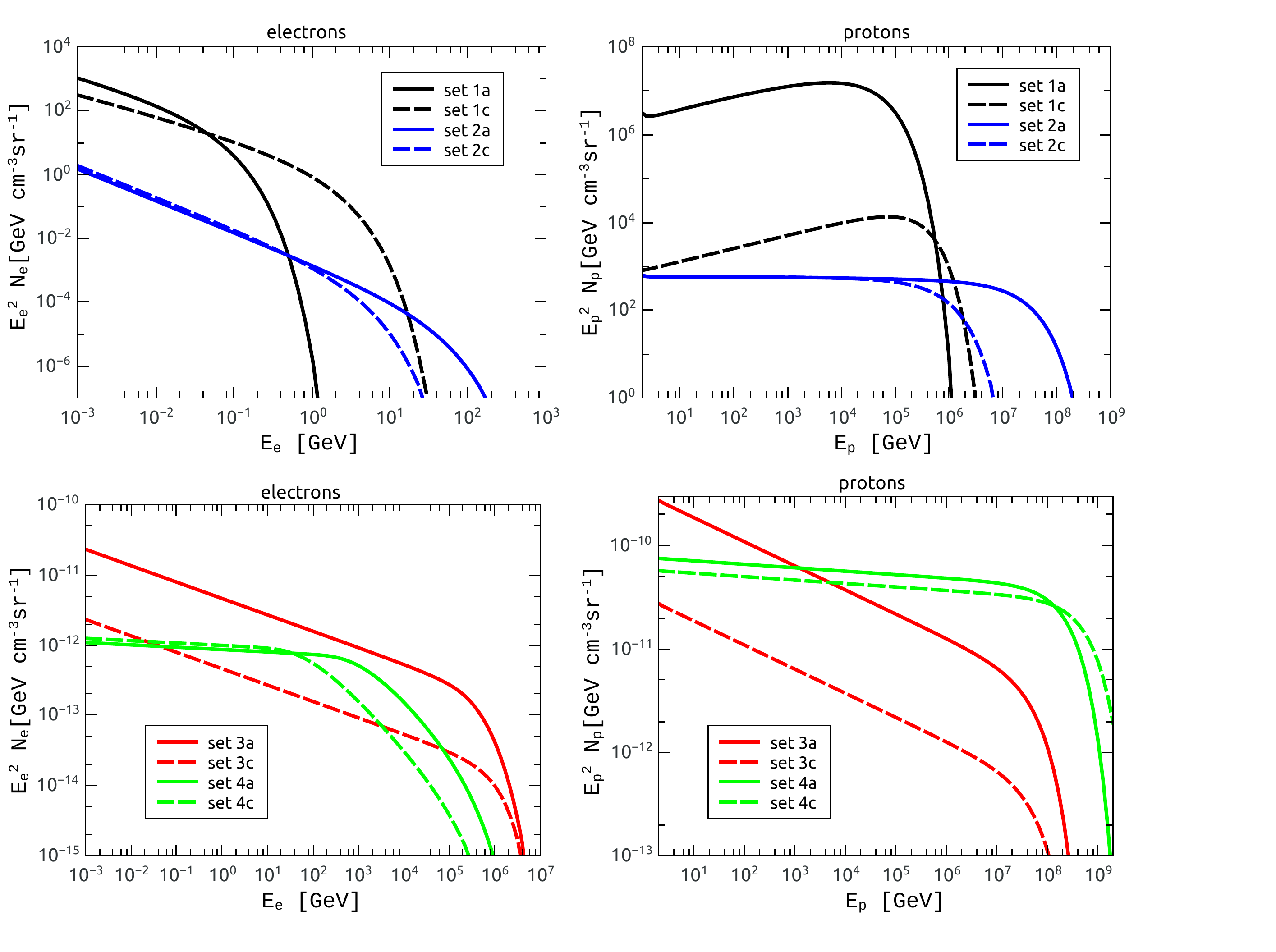}
\caption{Electron and proton distributions for different parameter sets. \label{fig:Nep}.}
\end{figure*}

First, we obtain the electron distribution ($N_e$) considering only synchrotron and adiabatic cooling, then we compute the SSC cooling rate and check that it is well below the total cooling rate adopted. Next, we compute the $p\gamma$ cooling rate and use it along with the other cooling processes for protons in order to obtain the proton distribution. The distributions $N_e$ and $N_p$ obtained for the different parameter sets are shown in Fig. \ref{fig:Nep}.

\begin{figure*}[tbp]
\includegraphics[width=.99\textwidth,trim=0 0 0 0,clip]{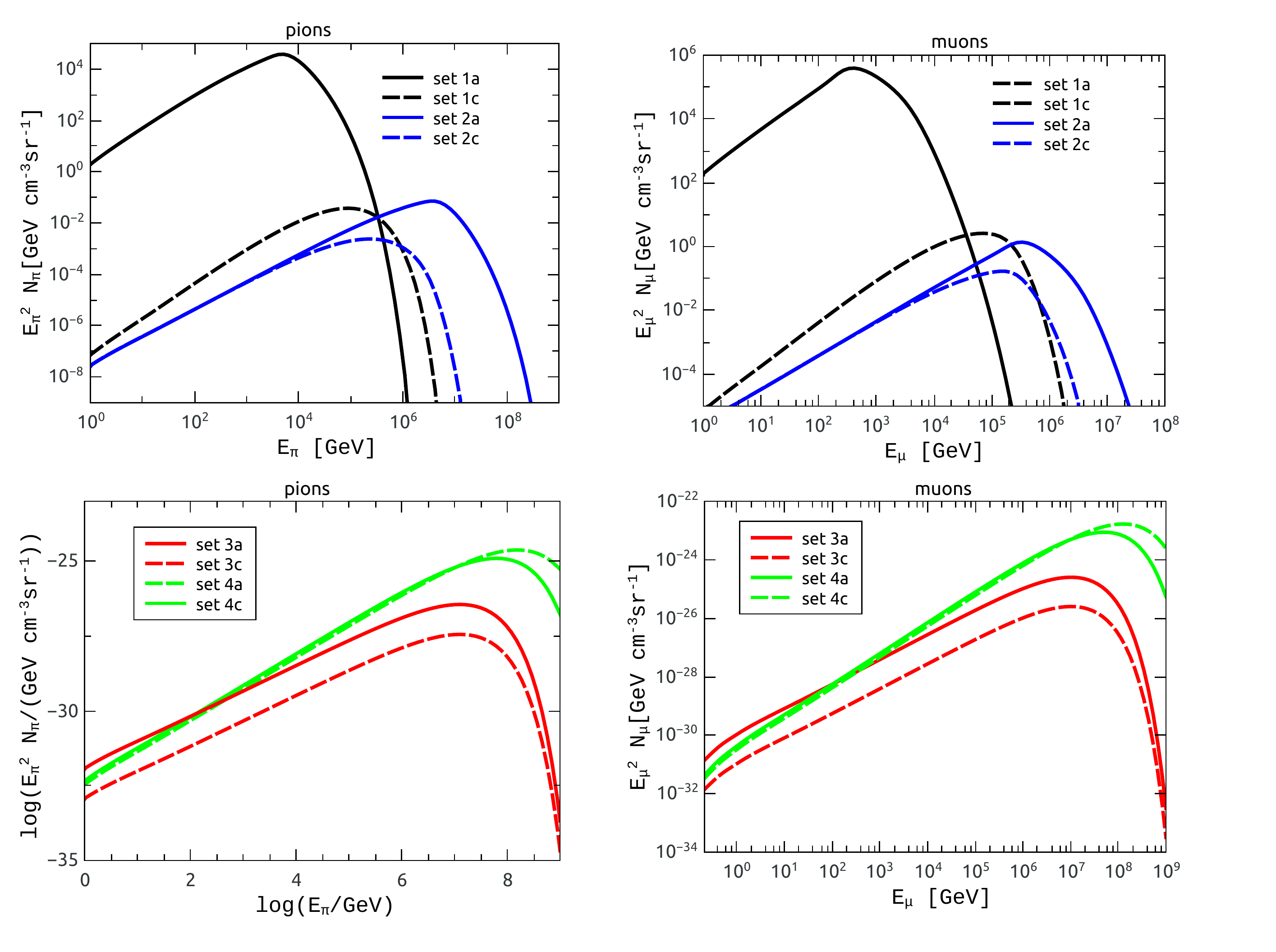}
\caption{Pion and muon distributions for different parameter sets. \label{fig:Npimu}}
\end{figure*}

{With the proton distribution, it is possible to obtain the source term of charged pions $Q_{\pi}$ which in this work are dominantly producuded by $pp$ interactions: 
}\begin{equation}
Q_{\pi}(E_\pi)=n_{\rm c} c\int_{0}^{1} \frac{dx}{x} N_p \left(\frac{E_\pi}{x}\right)
F_\pi\left(x,\frac{E_\pi}{x}\right)\sigma_{pp}^{\rm(inel)} \left(\frac{E_\pi}{x}\right).
\label{Qpipp}
\end{equation}
{Here, the $pp$ cross section $\sigma_{pp}^{\rm(inel)}$  and the distribution of pions produced per $pp$ collision $F_\pi$ are given in Ref. \citep{kelner2006}. The density of cold protons is taken as $n_{\rm c}\approx \rho_k/(m_p c^2)$ for the zones placed at the inner jet, and it is fixed as $n_{\rm c}\approx 4{\rm cm^{-3}}$ if the zone is located at the jet termination region \citep{bordas2009}.}
%
{This allows us to solve a transport equation like Eq. (\ref{equ}), but adding the decay term $N_\pi/ T_{\rm dec}$ on the left member. In turn, making use of the obtained pion distributions, we can obtain a source term for left-handed muons as \citep{lipari2007}}
\beq
  Q_{\mu^-_L,\mu^+_R}(E_\mu)= \int_{E_\mu}^{\infty} dE_\pi 
  \frac{N_{\pi}(E_\pi)}{T_{\pi,{\rm dec}}(E_\pi)} \frac{dn_{\pi^- \rightarrow
  \mu^-_L}}{dE_\mu}(E_\mu;E_\pi), \label{QmuL}
 \eeq  
and for right-handed muons:
\beq
  Q_{\mu^-_R,\mu^+_L}(E_\mu)= \int_{E_\mu}^{\infty} dE_\pi 
  \frac{N_{\pi}(E_\pi)}{T_{\pi,
{\rm dec}}(E_\pi)} \frac{dn_{\pi^- \rightarrow
  \mu^-_R}}{dE_\mu}(E_\mu;E_\pi). \label{QmuR}
\eeq  
{ Similarly as for pions, a transport equation is solved for muons. For illustration, we show the resulting distributions $N_\pi$ and $N_\mu$ in Fig. \ref{fig:Npimu} for different sets of parameters. We note that the use of the transport equation for pions and muons allows to take into account  the sychrotron losses that they can undergo in the case of high magnetic fields as are suppposed to be present at the inner jets \citep{magnetic2009}. We have not considered the possibility of pion and muon acceleration, since we are working within a one-zone model where the particles enter the zone already accelerated. More detailed studies would be needed in order to include the details of the particular acceleration mechanism, and in particular, the relation between the escape rate from the acceleration zone into the cooling zone and the acceleration rate itself \citep{twozone2014,klein2013,winter2014}.}


\section{Photon and neutrino output}
In this section, we show our results for the electromagnetic and neutrino emission that can be expected to arrive on Earth in the case that populations of relativistic electrons and protons can be accelerated in the jets of SS433, with the physical conditions described by the parameters of Table \ref{table.params}.

\subsection{Electromagnetic emission}
The most significant processes for the production of electromagnetic emission within the model discussed here are synchrotron radiation by electrons at the lower energies and proton-proton interactions at higher energies. The former process implies an emissivity, i.e. $({\rm number \ of \ photons})/ ({\rm  cm^{3} \, s  \, sr})$, which is given by
\be
Q_{\gamma, \, \rm syn}(E_\gamma')=\frac{\sqrt{2}e^3 B}{m_e c^2 h}\frac{4\pi}{E_{\rm cr}} \int_{m_e c^2}^\infty dE'\int_{\frac{E'_\gamma}{E_{\rm cr}}}^\infty d\zeta K_{\frac{5}{3}}(\zeta)N_e(E_e),
\ee
where $K_{\frac{5}{3}}$ is the modified Bessel function of order $\frac{5}{3}$, and the critical photon energy is
$$E_{\rm cr}= \frac{\sqrt{6}h\, e\, B}{4\pi m_e c}\left(\frac{E'}{m_e c^2}\right)^2.$$

The emissivity of photons due to $pp$ interactions is
\be
Q_{\gamma, \, pp}(E'_\gamma)= n_c c \int_0^1\frac{dx}{x}
N_p(E'_\gamma/x)F_{pp\rightarrow\gamma}(x,E'_\gamma/x) \sigma_{pp}^{\rm (inel)}(E'_\gamma/x),
\ee
where $F_{pp\rightarrow\gamma}(x,E_p)$ is a fitting function as defined in Ref. \citep{kelner2006}, and $n_{\rm c}$ is the density of cold protons.
For the sets of parameters considered, these processes turn out to be dominant over other emission mechanisms, such as SSC, proton-photon, and proton synchrotron, which can be safely neglected when computing the broadband spectral energy distribution of photons (SED).

\begin{figure}[tbp]
\includegraphics[width=.52\textwidth,trim=0 1 50 1,clip]{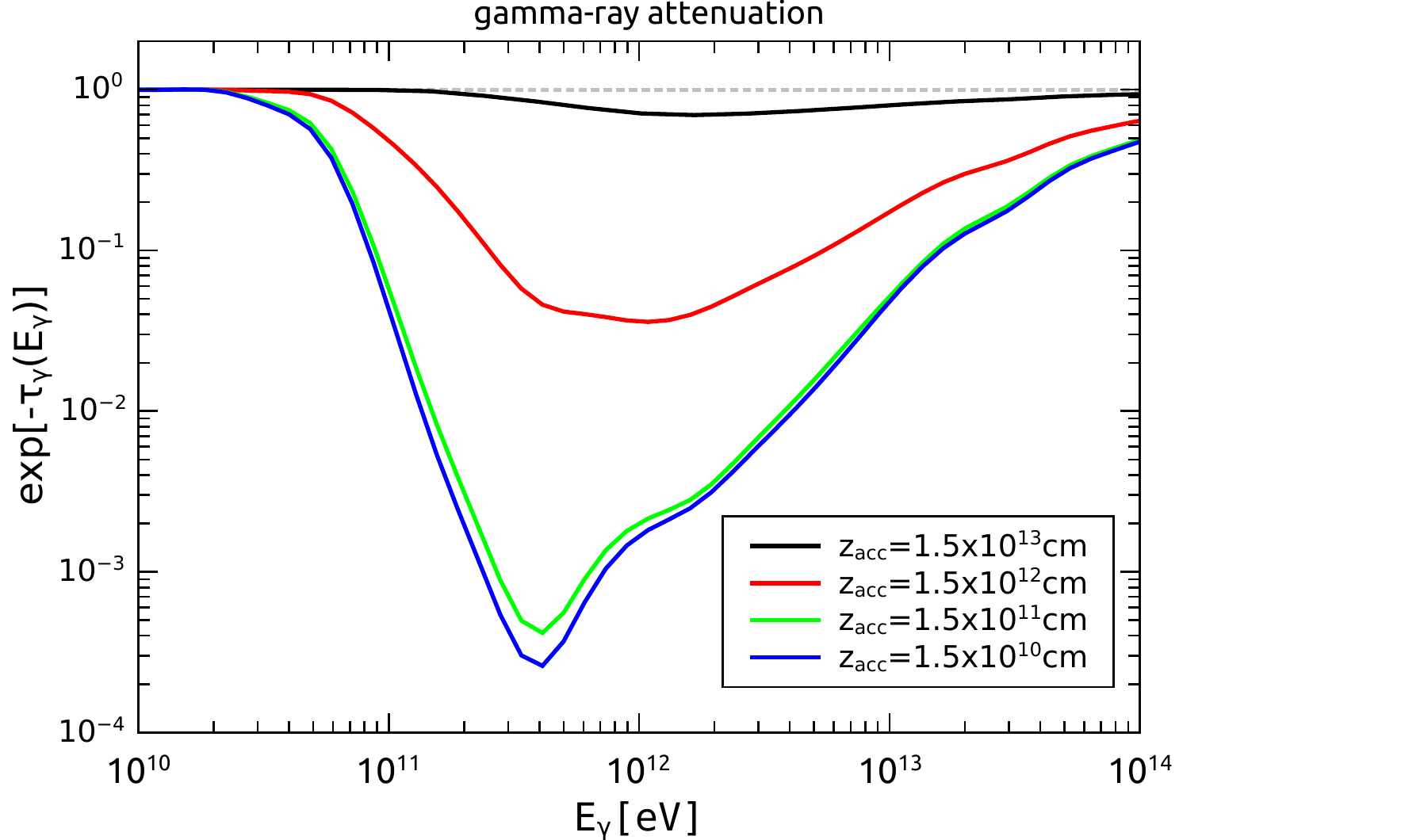}
\caption{Gamma-ray attenuation factor for different positions $z_{\rm acc}$ along the jets. \label{fig:attenuation}}
\end{figure}

In order to account for the emission that would arrive on Earth, it is necessary to include the effect of photon absorption, {which is significant only for the cases of production at the inner jet}.
At high energies ($E_\gamma\gtrsim 1 \, {\rm GeV}$), gamma rays can be efficiently absorbed by $\gamma \gamma$ interactions with lower energy photons to create $e^+e^-$ pairs. Internal absorption, where the target photons are due to the synchrotron emission from the electrons in the jet, is accounted for with an optical depth given by
\be
\tau_\gamma^{\rm(int)} = R_{\rm jet} \int_{E_{\rm ph \, min}}^\infty dE_{\rm ph} {n_{\rm ph}(E_{\rm ph})} \sigma_{\gamma\gamma}(E_\gamma,E_{\rm ph}).  
\ee
External absorption of gamma rays occurs due to the presence of starlight photons and UV and IR photons emitted from the disk. Additionally, $\gamma N$ interactions with nucleons that constitute the disk itself can cause important absorption if the production takes place very close to the jet base. These effects were studied in detail in previous works \citep{reynoso2008a,reynoso2008b}, and it is found that if gamma rays are emitted at distances $z_{\rm acc}\lesssim 10^{12}{\rm cm}$ from the BH, absorption can imprint a modulation in the observed flux consistent with the precession of the jets and disk. For emission at longer distances from the BH, no periodic absorption features would be produced. In the present work, we consider the orientation of the jet corresponding to a precessional phase $\psi=0$, i.e., the situation when the jet is pointing as close as possible to the line of sight, making an angle of $i_{\rm jet}=57^\circ$ with it. For this particular orientation, the absorption effects are minimized, and hence we can interpret our results for the photon flux on Earth as an upper bound, since it will be reduced by a higher absorption at other precessional phases. We then consider the total optical depth for $\gamma\gamma$ and $\gamma N$ absorption $\tau_{\gamma}(E_\gamma)$ corresponding to the mentioned configuration and taken from Ref. \citep{reynoso2008b}. For illustration, we plot in Fig. \ref{fig:attenuation} the attenuation factor $\exp[-\tau_{\gamma}(E_\gamma)]$ corresponding to different positions $z_{\rm acc}$ in the jet.

At lower energies ($E_\gamma \sim (1-10^3){\ \rm eV}$), photons can get absorbed by photoionization along the column of neutral gas along the line of sight directed towards the inner jet, $N_H\approx 3\times 10^{22}{\rm cm^{-2}}$ (e.g. \citep{marshall2013}). The corresponding optical depth is $\tau_{\gamma H}=N_H\sigma_{\gamma H}(E_\gamma)$, where $\sigma_{\gamma H}(E_\gamma)$ is the photoionization cross section for galactic abundances of neutral hydrogen and dust following Ref. \citep{ryter1996}.

\begin{figure*}[t]
\includegraphics[width=0.93\textwidth,trim=0 0 0 0,clip]{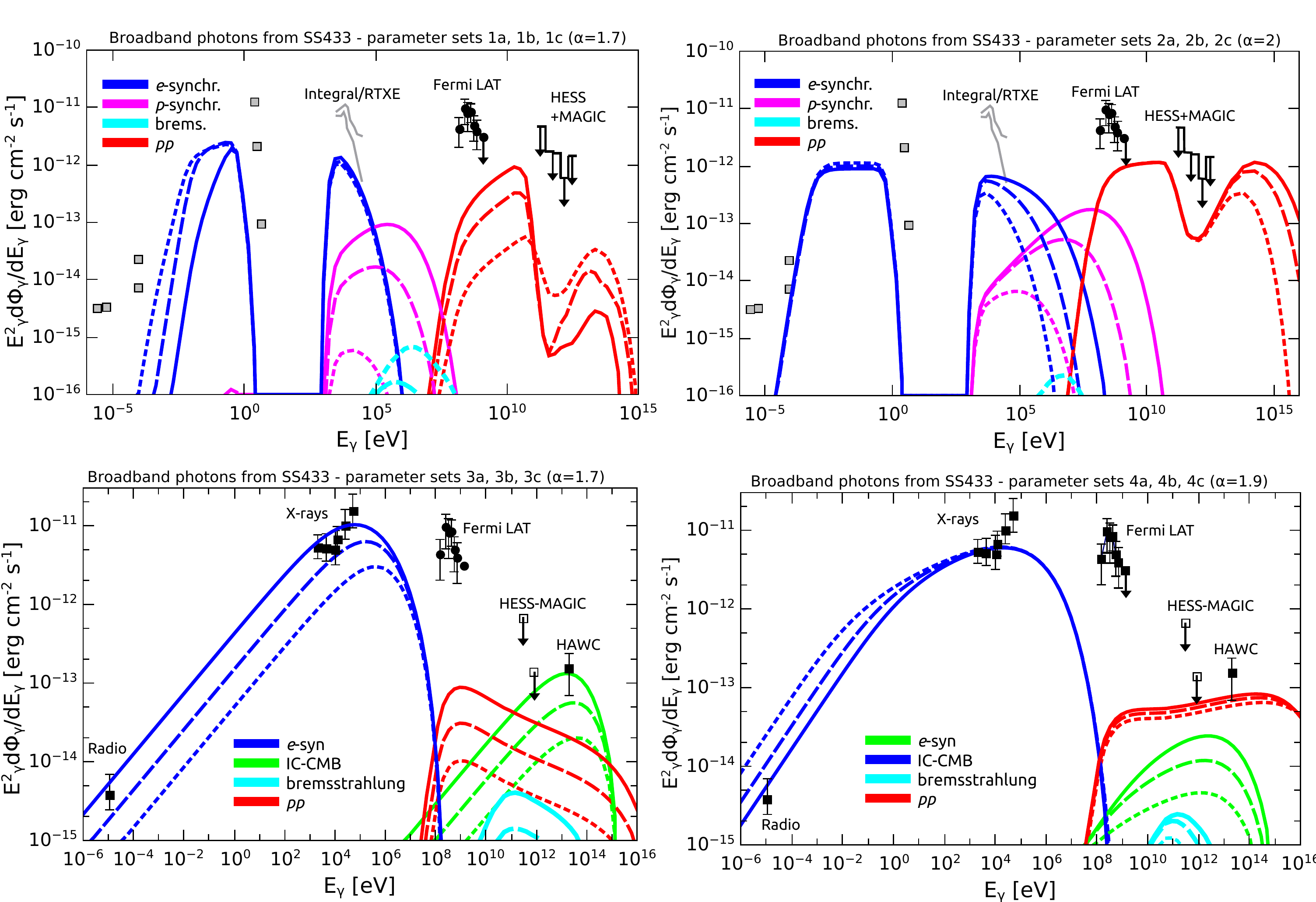}
\caption{Photon fluxes obtained for different parameter sets \label{fig:photonfluxes}.}
\end{figure*}

Including the effects of absorption, the differential photon flux to arrive on Earth at a distance $d=5.5\, {\rm kpc}$ can be computed as
\be
  \frac{d\Phi_\gamma}{dE_\gamma}= \frac{D\,\Delta V}{d^2}Q_\gamma(E'_\gamma)\exp\left[{-\tau_{\gamma}(E'_\gamma)}\right],
\ee
where the observed photon energy $E_\gamma$ is related to the comoving one as $E_\gamma= D E'_\gamma$, with the Doppler factor given by $D=(\Gamma - \Gamma\beta\cos{i_{\rm jet}} )^{-1}$.
We show the obtained results in Fig. \ref{fig:photonfluxes} for the different sets of parameters used (\ref{table.params}). There we include observational data taken, for reference, from the multi-wavelength campaign in Ref. \citep{chakra2005} for energies $E_\gamma< 10^5{\rm eV}$ {in the case of emission from the inner jet}. Gamma-ray data corresponds to the positive detection by Fermi LAT for $E_\gamma \in (0.1-1) \, {\rm GeV}$ \citep{bordas2015} and to the upper limits obtained by the joint HESS+MAGIC observations at $E_\gamma \in (200-5000) \ {\rm GeV}$ \citep{hessmagic2017}. {In the case of emission from the termination region of the jet, we adopted the same data set that were used by the HAWC collaboration in Ref. \citep{hawc2018}.}
We note that for the parameter sets 1's and 2's, the correction by absorption is actually overestimated since the size of the emission zone is large ($\sim 10^{13}{\rm}$), and we here consider the value of the optical depth at $z_{\rm acc}\sim 10^{12}{\rm cm}$. However, here we are interested in obtaining the maximum allowed neutrino fluxes, and hence this approximation only implies that our results for such maximum fluxes of gamma rays are actually conservative, i.e., they could be lower if a detailed treatment of gamma-ray absorption was applied.

\subsection{Neutrino emission}
The relevant neutrino emission process in the cases studied here are $pp$ production of pions, which decay to neutrinos and muons. The latter, in turn, decay giving more neutrinos.
The emissivity $\nu_\mu+\bar{\nu}_\mu$ from direct pion decays can be obtained following Ref. \citep{lipari2007}:
\be
Q_{\pi\rightarrow\nu_\mu}(E_\nu)= \int_{E}^{\infty}dE_\pi
T^{-1}_{\pi,\rm d}(E_\pi)N_\pi(E_\pi)
\frac{H(1-r_\pi-x)}{E_\pi(1-r_\pi)},
 \ee
where $x=E_\nu/E_\pi$ and typical timescale of pion decay is $T_{\pi,\rm d}=2.6 \frac{E_\pi}{m_\pi c^2}\times 10^{-8}{\rm s}$.
The contribution from muon decays to $\nu_\mu+\bar{\nu}_\mu$ is
\begin{multline}
Q_{\mu\rightarrow\nu_\mu}(E_\nu)= \sum_{i=1}^4\int_{E}^{\infty}\frac{dE_\mu}{E_\mu} T^{-1}_{\mu,\rm
d}(E_\mu)N_{\mu_i}(E_\mu) \\ \times \left[\frac{5}{3}-
3x^2+\frac{4}{3}x^3 +\left(3x^2-\frac{1}{3}-\frac{8x^3}{3}\right)h_{i}
\right],
\end{multline}
where $x=E_\nu/E_\mu$, $\mu_{1,2}=\mu^{-,+}_L$, the muon lifetime is $T_{\mu,\rm d}=2.2\frac{E_\mu}{m_\mu c^2}\times 10^{-6}{\rm s}$, and
$\mu_{3,4}=\mu^{-,+}_R$. The helicity of the muons is $h=1$ for right-handed and $h=-1$ for left- handed muons. 
As for $\nu_e+\bar{\nu}_e$ , the emissivity from the decay of muons is given by
\begin{multline}
Q_{\mu\rightarrow\nu_e}(E_\nu)= \sum_{i=1}^4\int_{E_\nu}^{\infty}\frac{dE_\mu}{E_\mu} T^{-1}_{\mu,\rm
d}(E_\mu)N_{\mu_i}(E_\mu,t) \\ \times \left[2-
6x^2+4x^3 +\left(2- 12x+ 18x^2-8x^3\right)h_{i}\right].
\end{multline}

Taking into account the effect of neutrino oscillation during their propagation to Earth, the differential muon neutrino and antineutrino flux of energy $E_\nu=D \,E'_\nu$ can be computed as
\be
\frac{d\Phi_{\nu_\mu}}{dE_\nu} = \frac{D\,\Delta V}{d^2}\left[Q_{\nu_\mu}(E'_\nu) \, P_{\nu_\mu\rightarrow \nu_\mu} +
Q_{\nu_e}(E'_\nu) \, P_{\nu_e\rightarrow \nu_\mu}\right],
\ee
where $P_{\nu_{\mu}\rightarrow \nu_{\mu}}\simeq 0.369$ is the probability that the generated $\nu_\mu$ or $\bar{\nu}_\mu$ remain of the same flavour, and $P_{\nu_{e}\rightarrow \nu_{\mu}}\simeq 0.255$ is the probability that electron neutrinos or antineutrinos can oscillate into muon neutrinos or antineutrinos. The specific numeric values of these probabilities are derived from the unitary mixing matrix $U_{\alpha j}$, which is fixed by three mixing angles: $\theta_{12}\simeq 34^\circ$, $\theta_{13}\simeq 9^\circ$, and $\theta_{23}\simeq 45^\circ$ \citep{gonzalezgarcia2012}.

\begin{figure}[t]
\includegraphics[width=.45\textwidth,trim=0 5 0 0,clip]{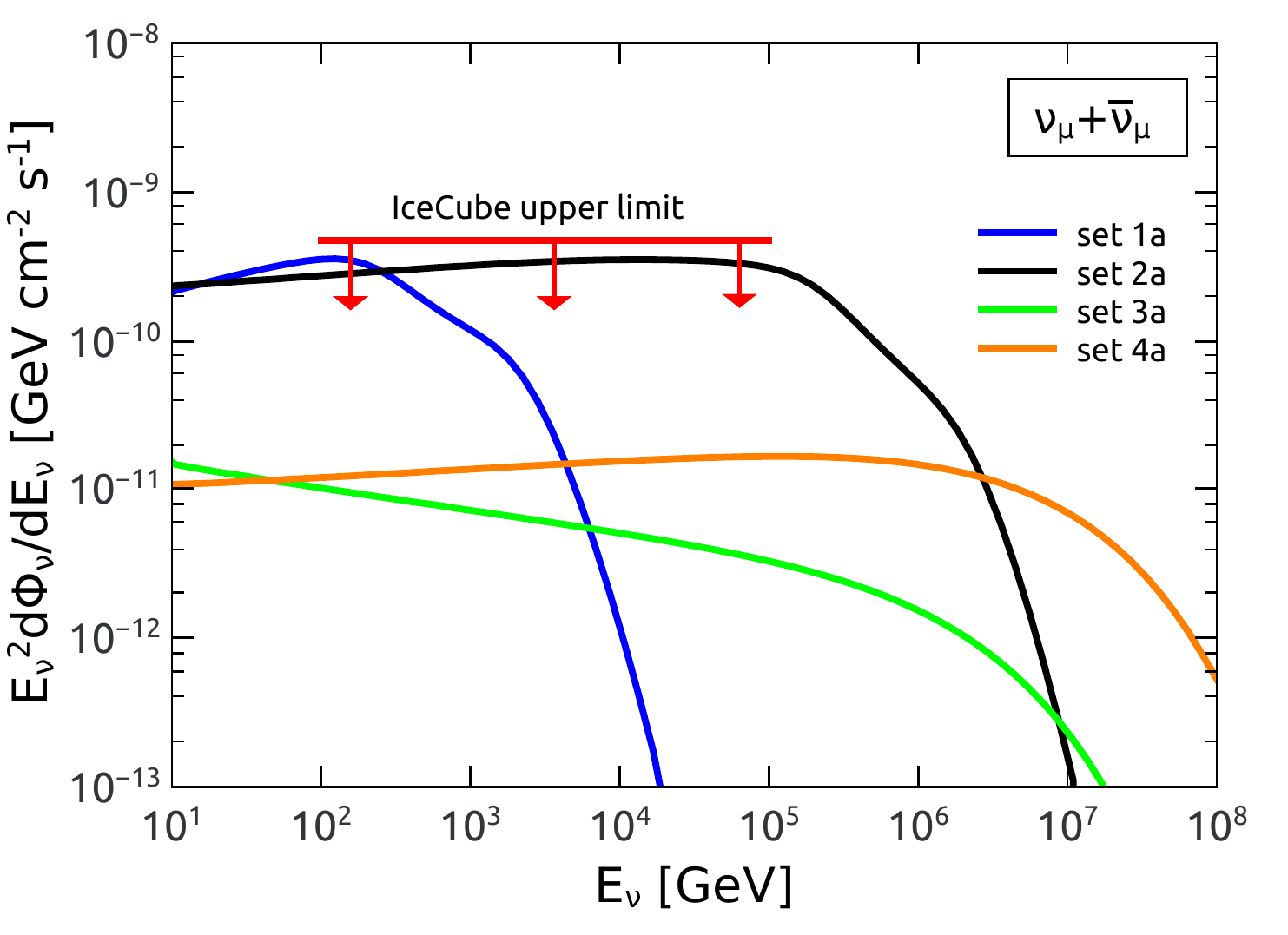}
\vspace{-4mm}
\caption{Muon neutrino and antineutrino fluxes for different parameter sets \label{fig:nufluxes}.}
\end{figure}

\section{Discussion}

In this work, we have explored different possibilities for neutrino production in the microquasar SS433 due to particle acceleration in the inner jets or at their termination sites, making use of a simple model applied to describe the injection and interaction of high energy electrons and protons. We have computed the radiative output in broadband photons and compared it with typical data available from radio to gamma rays for three different cases of injected spectrum of primary particles $Q_{e,i}\propto E^{-\alpha}$, with $\alpha=\{1.7, 1.9, 2\}$. For each of these situations, the electromagnetic emission was checked not to be in conflict with observations, and at the same time, the relativistic proton content was pushed as high as possible without surpassing the neutrino upper limit. 
 {We have considered six combinations of parameters for emission from the inner jet (sets 1a, 1b, 1c, 2a, 2b, 2c), and other six combinations for emission at the termination region (3a, 3b, 3c, 4a, 4b, 4c).} {The sets of parameters used were chosen in order to understand the consequences in the emission of a change in a given parameter. For instance, the parameter sets 1a, 1b, and 1c differ only by the value of $q_m$ which is taken as $0.1$, $0.01$, and $0.001$, respectively. It can be seen that as the value of $q_m$ decreases, the position of injection along the inner jet $z_{\rm acc}$ is placed at greater distances from the BH. This implies a less efficient emission, but at the same time, less $\gamma-\gamma$ absorption. The parameter sets 2a, 2b, and 2c differ only by the value of the acceleration efficiency $\eta$, which is taken as $10^{-2}$, $10^{-3}$, and $10^{-4}$, respectively. The effect of the increase of $\eta$ is an increase in the maximum energy of the primary particles, and hence of the gamma-rays and neutrinos. }
{In the cases of emission from the terminal region of the jet, the parameter sets 3a, 3b, and 3c differ only by the value of the diffusion coefficient $D_0$: $3\times 10^{27}{\rm cm^2s^{1}}$, $10^{28}{\rm cm^2s^{1}}$, and $3\times 10^{28}{\rm cm^2s^{1}}$, respectively. As this increases, particles escape faster from the production zone, and emission is reduced. We note that according to these three sets, the $25$TeV flux detected by HAWC is explained basically by IC interactions on the CMB \citep{hawc2018}. The only possibilities for this emission to be due to $pp$ interactions arise if the diffusion coefficient was smaller than the typical values, for instance, $D_0\sim 5\times 10^{26}$, as is adopted in the sets 4a, 4b, and 4c. In these sets, the magnetic field has values of $4\times 10^{-5}{\rm G}$, $6\times 10^{-5}{\rm G}$, and $10^{-4}{\rm G}$, respectively. } 
  
{Overall, among the various possibilities explored, the ones that give the maximum neutrino output are 1a, 2a, 3a, and 4a, which are shown in Fig. \ref{fig:nufluxes}, as compared to the latest IceCube upper limit.}

  In the typical case of a proton injection of $E_p^{-2}$ (sets 2a, 2b, 2c), we have chosen an injection point $z_{\rm acc}=1.5 \times 10^{12}\, {\rm cm}$ that implies a moderate gamma-ray absorption effect for $E_\gamma>100 \, {\rm GeV}$, and this fact places the emission below HESS+MAGIC upper bounds (see Fig. \ref{fig:photonfluxes}, top-right panel) for a corresponding neutrino flux barely allowed by the IceCube limit (see  Fig. \ref{fig:nufluxes}, black curve).
  For higher values of $z_{\rm acc}$, absorption would not be so efficient, and hence the intrinsic production of gamma rays should be lower, leading to a correspondingly lower neutrino emission.
  Considering the prospects for the evolution of the sensitivity expected for IceCube versus time of operation, in ten years more of livetime, their minimum detectable flux would be reduced by a factor of two, roughly (see Ref. \citep{icecube2017}, Fig. 15).
Hence, the possibilities to observe neutrinos corresponding to $\alpha \approx 2$ are that neutrinos could be produced at a detectable level only at a position $z_{\rm acc} \lesssim 10^{13}{\rm cm}$ in the jet, otherwise, negligible absorption of VHE gamma rays would imply a flux above the constraints by HESS+MAGIC. {This applies also to the production at the terminal regions of the jets, which should be, at most, at the level of the HAWC detection in $25{\,\rm TeV}$ gamma rays, and below the upper limits by HESS+MAGIC at somewhat lower energies. Co-produced neutrinos at these locations of the system must be, at most, at the level indicated by the set 4a curve in Fig. \ref{fig:nufluxes}, i.e., beyond the reach of IceCube.}
Conversely, in the case of a future positive detection of VHE gamma-rays and no associated detectable neutrino counterpart, this would imply a leptonic origin of such radiation.
Further observations of the VHE gamma rays with forthcoming instruments such as CTA \cite{cta2017} and IceCube-Gen2 \citep{icecubegen22015} would help to assess the plausibility of such possibilities.

\section*{Acknowledgments}
 We thank CONICET (PIP-2013-2015 GI 160) and Universidad Nacional de Mar del Plata for financial support.


\begin{thebibliography}{}
\bibitem{mirabel1998} I. F. Mirabel,  L. F. Rodriguez, Nature {392} (1998) 673
\bibitem{bosch2008} V. Bosch-Ramon, D. Khangulyan,
  Int.\ J.\ Mod.\ Phys.\ D {18} (2008) 347
 \bibitem{fabrika2006}
   S. Fabrika, 
  Comments Astrophys.\ Space Phys.\  {12} (2004) 1
\bibitem{fuchs2006} Y. Fuchs, L. K. Miramond, P. \'{A}brah\'{a}m, Astron. Astrophys. {445} (2006) 1041
\bibitem{gies2002} D. R. Gies, M. V. McSwain, R. L. Riddle, Z. Wang, P. J. Wiita, D. W. Wingert, Astrophys. J. {566} (2002) 1069
\bibitem{migliari2002}S. Migliari, R. Fender, M. M\'endez, Science {297} (2002) 1673
\bibitem{dubner1998}  G. M. Dubner, M. Holdaway, W. M. Goss, I. F. Mirabel, AJ {116} (1998) 1842
\bibitem{jeffrey2016} R.~M. Jeffrey, K. M. Blundell, S. A. Trushkin, A. J. Mioduszewski, 
  Mon. Not. R. Astron. Soc.   {461} (2016)  312
\bibitem{chere2005} A.~M. Cherepashchuk  { et al.}, 
  Astron. Astrophys.  {437} (2005) 561
\bibitem{bordas2015}P. Bordas, R. Yang, E. Kafexhiu, F. Aharonian,
  Astrophys. J. {807} (2015)  L8  
\bibitem{xing2018} 
  Y.~Xing, Z.~Wang, X.~Zhang, Y.~Chen and V.~Jithesh,
  arXiv:1811.09495 [astro-ph.HE].  
\bibitem{hessmagic2017}
 M.~L. Ahnen {et al.} [MAGIC and H.E.S.S. Collaborations]
  Astron. Astrophys. (2017) 
\bibitem{veritas2017}
 P. Kar, [VERITAS Collaboration],
  arXiv:1708.04967 [astro-ph.HE].  
  
\bibitem{hawc2018} A.~U. Abeysekara, et al., Nature 562 (2018) 82 
  
\bibitem{reynoso2008b}
 M. M. Reynoso, G. E.Romero, H. R. Christiansen, Mon. Not. R. Astron. Soc. {387} (2008) 1745
\bibitem{icecube2017}
  M.~G. Aartsen {et al.} [IceCube Collaboration], 
  Astrophys. J.  {835} (2017) 151
\bibitem{romero2003}
 G. E. Romero, D. F. Torres, M. M.,  Kaufman-Bernad\'{o}, I. F.  Mirabel, Astron. Astrophys. {410} (2003) L1
\bibitem{romerovila2008} G. E. Romero, G.~S. Vila, 
  A\&A  {485} (2008) 623  
\bibitem{magnetic2009}
 M. M. Reynoso, G. E. Romero, .
 A\&A  {493} (2009) 1
\bibitem{vilaromero2010} G. S. Vila,  G. E. Romero, Mon. Not. R. Astron. Soc. 403 (2010) 1457    
\bibitem{krolik1999}
 J. H. Krolik, Active galactic nuclei : from the central black hole to the
galactic environment, ed. Princeton University Press (1999) Princeton
\bibitem{gaisser1991}
 T. K. Gaisser, Cosmic Rays and Particle Physics, pp. 295. ISBN 0521326 672. Cambridge, UK (1991) Cambridge University Press.
\bibitem{rieger2007}
  F.~M.~Rieger, V.~Bosch-Ramon, P.~Duffy,
  Astrophys.\ Space Sci.\  {\bf 309} (2007) 119    
\bibitem{begelman1990}M. C. Begelman, B. Rudak, M. Sikora,  Astrophys. J. 362 (1990) 38
\bibitem{hillas1984}A. M. Hillas, ARAA 22 (1984) 425
\bibitem{bordas2009}P.  Bordas, V.  Bosch-Ramon, J. M.  Paredes, M.  Perucho
Astron. Astrophys. 497 (2009) 325
DOI: 10.1051/0004-6361/200810781
\bibitem{venya1990} V.~S. Berezinskii, S.~V. Bulanov, V.~A. Dogiel, \& V.~S. Ptuskin. Amsterdam: North-Holland (1990) edited by Ginzburg, V.L.
\bibitem{ormes1988} J.~F. Ormes, M.~E. Ozel, D.~J. Morris,  Astrophys. J. 334 1988) 722 

\bibitem{leptohadronic2011}
 M.~M Reynoso, M. C. Medina, G. E. Romero, A\&A 531 (2011) A30
  \bibitem{lipari2007}
 P. Lipari, M. Lusignoli, D. Meloni, 
  Phys.\ Rev.\ D {75} (2007) 123005

\bibitem{twozone2014} 
  M.~M.~Reynoso,
  Astron.\ Astrophys.\  {564}, A74 (2014)  
\bibitem{klein2013} 
  S.~R.~Klein, R.~E.~Mikkelsen and J.~Becker Tjus,
  Astrophys.\ J.\  {779}, 106 (2013)
\bibitem{winter2014} 
  W.~Winter, J.~Becker Tjus and S.~R.~Klein,
  Astron.\ Astrophys.\  {569}, A58 (2014)  

\bibitem{kelner2006}
  S.~R. Kelner, F.~A. Aharonian, V.~V. Bugayov, 
  Phys.\ Rev.\ D {74} (2006) 034018
   Erratum: [Phys.\ Rev.\ D {79} (2009) 039901]
\bibitem{reynoso2008a}
 M. M. Reynoso, G. E. Romero, H. R. Christiansen, Astropart. Phys {28} (2008) 565
\bibitem{marshall2013}
  H.~L.~Marshall, C.~R.~Canizares, T.~Hillwig, A.~Mioduszewski, M.~Rupen, N.~S.~Schulz, M.~Nowak, S.~Heinz,
  Astrophys.\ J.\  {775} (2013) 75 

\bibitem{ryter1996}
 C. E. Ryter,  Ap{\&}SS {236} (1996) 285
\bibitem{chakra2005}
  S.~K. Chakrabarti  {\it et al.} 
  Mon. Not. R. Astron. Soc. {362} (2005) 957   


\bibitem{gonzalezgarcia2012} 
 M. C. Gonzalez-Garc\'{\i}a, M. Maltoni, J. Salvado, T. Schwetz, 
   JHEP  {1212} (2012) 123 
\bibitem{cta2017} 
  B.~S.~Acharya {et al.} [Cherenkov Telescope Array Consortium],
  arXiv:1709.07997 [astro-ph.IM]

\bibitem{icecubegen22015} 
  M.~G.~Aartsen {\it et al.} [IceCube Collaboration],
  arXiv:1510.05228 [astro-ph.IM].
   		
  

 \end{thebibliography}
\end{document}